# Lattice models of non-trivial "optical spaces" based on metamaterial waveguides

Alexei I. Smolyaninov*, Igor I. Smolyaninov

*Department of Electrical and Computer Engineering, University of Maryland, College Park, MD 20742, USA*
*Corresponding author: a.smoly@gmail.com*

Metamaterials are being used to model various exotic "optical spaces" for such applications as novel lenses and cloaking. While most effort is directed towards engineering of continuously changing dielectric permittivity and magnetic permeability tensors, an alternative approach may be based on lattices of metamaterial waveguides. Here we demonstrate the power of the latter technique by presenting metamaterial lattice models of various 4D spaces.

Electromagnetic metamaterials can be used to emulate highly unusual "optical spaces" enabling such applications as novel transformation optics (TO) based lenses and cloaking [1-3]. This development is enabled by the newfound freedom of continuous control of the local dielectric permittivity $\varepsilon_{ik}$ and magnetic permeability $\mu_{ik}$ tensors in electromagnetic metamaterials. Even though less prominent, another direction in metamaterial research may be characterized as "lattice-based metamaterial models". These models are based on the networks of metamaterial waveguides, which control propagation of the electromagnetic signals within a given 3D volume. Examples of this approach may be found in couple of recent experimental realizations of electromagnetic cloaking [4,5]. The goal of our paper is to demonstrate that the latter approach to "optical space" design with metamaterials is also very powerful, and it may supplement the more common approach of continuous engineering of $\varepsilon_{ik}$ and $\mu_{ik}$ tensors. For example, lattice-based metamaterial designs allow straightforward realization of various non-Euclidean "optical spaces", which were suggested as a pathway to broadband cloaking [6].

We demonstrate the ultimate power of lattice-based approach by presenting metamaterial lattice models of various hypercubic 4D spaces, which are impossible to emulate using conventional transformation optics. While our claim may seem unusual, we must point out that "hypercubic network architectures" for multiprocessor computing are being widely described in computer engineering literature. It is well known that hypercubic lattice networks of any spatial dimension D can be projected onto 3D space. Moreover, many of these hypercubic networks had been realized in the experiment [7]. An obvious advantage of implementing a lattice-based 4D metamaterial "optical hyperspace" is that any 3D non-Euclidean space may easily fit into such 4D space [8]. Therefore, a natural platform for the proposed non-Euclidean broadband TO designs will be provided.

The simplest model of a 4D lattice projected onto regular 3D space is presented in Fig.1, which shows a perspective projection of an elementary unit of a 4D hypercubic spatial lattice, and a perspective projection of 2x2x2x2 region of the hypercubic lattice onto a region of 3D space. The elementary unit is filled with metal, except for the black lines, which represent pieces of thin single mode coaxial waveguides, which are engineered to have the same impedance

$$Z = \frac{1}{2\pi}\sqrt{\frac{\mu}{\varepsilon}}\ln\frac{D}{d} \qquad (1)$$

and optical length $L=a(\varepsilon\mu)^{1/2}$ (where $a$ is the waveguide length, $d$ is the diameter of the inner conductor, and $D$ is the inner diameter of the outer electrode). Connection of the cables in the vertexes is realized using beam splitters. Eight cubic faces of this lattice act as boundaries of the eight neighboring hypercubic cells (similar to the way in which each square face in a 3D cubic lattice acts as a boundary of the neighboring cube). Thus, by keeping $L$ and $Z$ constant for each edge in the lattice, we can create a good lattice model of the 4D space: electromagnetic signals launched into the lattice would travel from vertex to vertex as if they propagate in a 4D hypercubic lattice. Due to metal filling, the cables communicate only at the vertexes.

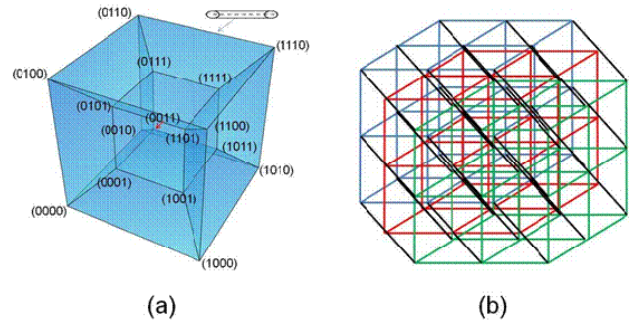

Fig. 1. (Color online) (a) Perspective projection of an elementary unit of a 4D hypercubic spatial lattice, which shows the (x,y,z,w) coordinates of each vertex. The elementary unit is filled with metal, except for the black lines, which represent pieces of thin single mode waveguides engineered to have the same impedance and the same optical length. (b) Perspective projection of a 2x2x2x2 region of the hypercubic lattice: three 2x2x2 elements of the cubic lattice shown in blue, red and green are shifted along the projected "fourth orthogonal direction" shown in black.

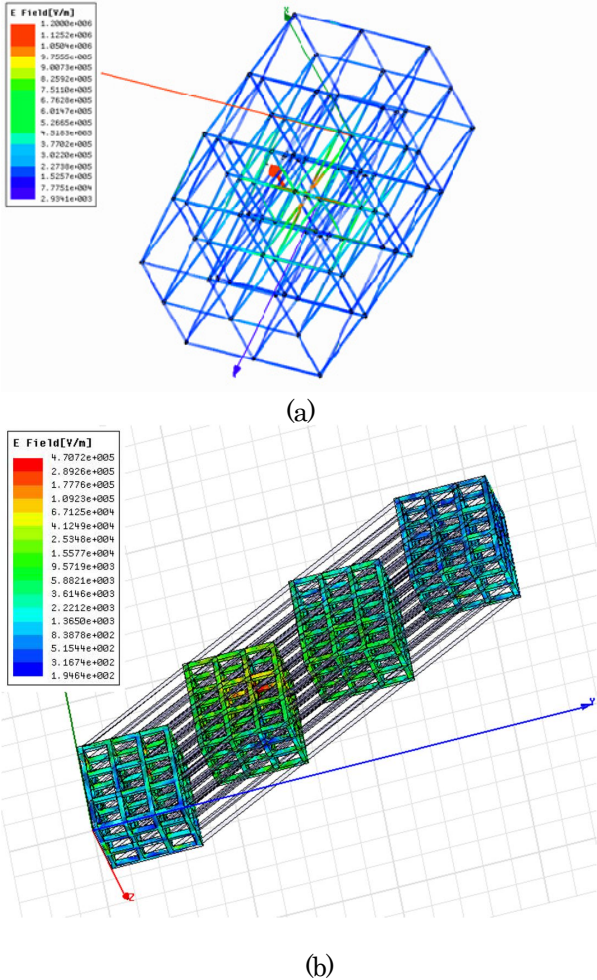

(a)

(b)

Fig. 2. (Color online) (a) HFSS simulations of electric field magnitude inside the 2x2x2x2 hypercubic lattice of waveguides. Radiation source is placed at the origin (0,0,0,0). (b) HFSS simulations of electric field magnitude inside the 3x3x3x3 hypercubic lattice of waveguides. For the sake of clarity, the field magnitude is shown only for four three-dimensional sections of the hypercube.

We have confirmed this approach by electromagnetic simulations performed using the Ansoft HFSS software package. Ansoft HFSS is an iterative matrix solver of Maxwell equations that applies the finite element method to a mesh generated for a physical model.. Our numerical data are presented in Figs.2 and 3. In Figure 2 we used the HFSS simulation engine to model a hypercubic lattice with the software suite's built-in Hertzian dipole source at the geometric center of the lattice. Both Fig 2a and Fig 2b used the same technique applied to a slightly different lattice structure, as shown in the figures. Maps of electric field distribution inside 2x2x2x2 and 3x3x3x3 hypercubic lattices of waveguides are presented in Figs. 2a and 2b, respectively. In both cases a radiation source was placed at the origin point (0,0,0,0) of the hypercubic lattice. The simulations were performed at 0.5THz. For the sake of clarity, in case of the 3x3x3x3 lattice the field magnitude is shown only for four three-dimensional sections of the hypercube (Fig.2b). The effective "four-dimensional distance" $R$ inside these lattices can be introduced as $R^2=x^2+y^2+z^2+w^2$, where $w$ is the distance along the projected "fourth spatial dimension". Measurements of field intensity $I$ as a function of $R$ are supposed to determine the effective dimensionality of the created lattice space. In the 3D space $I\sim 1/R^2$, while in the 4D space $I\sim 1/R^3$. Data plotted in Fig.3 came directly from the results shown in Figure 2b by taking the magnitude of the E field at the lattice points. As shown in Fig.3, field intensity measured at the lattice vertices plotted as a function of $R^{-3}$ is consistent with the four-dimensional character of electromagnetic field propagation inside the lattice (at small $R$ minor deviation from $R^{-3}$ behavior is natural due to near-field effects). These simulations demonstrate practical feasibility of creating metamaterial models of the 4D space. In fact, similar technique is also applicable to lattice models of other higher dimensional $D>4$ spaces, since higher dimensional lattices can be projected onto 3D space in a similar manner [7]. However, we must emphasize that the lattice models used in such applications as cloaking must be large enough to emulate "continuous" space [4,5].

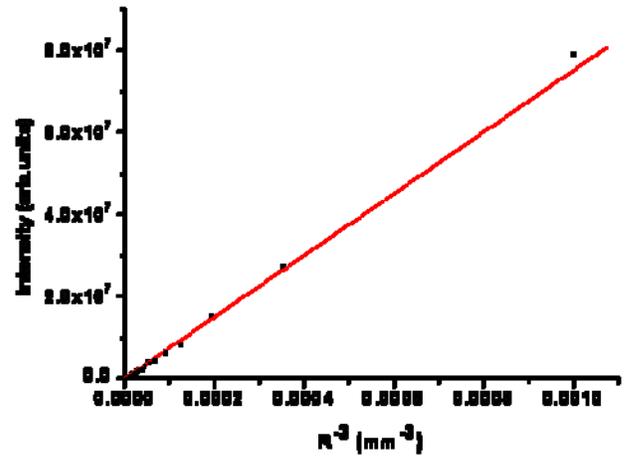

Fig. 3. (Color online) Field intensity measured at the lattice vertices plotted as a function of $R^{-3}$ is consistent with the four-dimensional character of field propagation at large $R$.

Even though making very large 4D lattice models may not be easy, a lattice model of the classic 5D Kaluza-Klein space-time [9]

$$ds^2 = g_{\alpha\beta}dx^\alpha dx^\beta + 2g_{\alpha 5}dx^\alpha d\phi + g_{55}d\phi^2 \qquad (2)$$

(where the Greek indices $\alpha=0, 1, 2, 3$ indicate coordinates of the 4D space-time, the fifth spatial dimension $\phi$ is compactified, and its circumference $g_{55}^{1/2}$ is small) seems to be feasible using $\varepsilon$ near zero waveguides described by Edwards et al. [10]. These waveguides would need to connect vertices of the emulated 4D hypercubic lattice with lattice coordinates $(x,y,z,0)$ and $(x,y,z,g_{55}^{1/2})$. As a result, spatial geometry of the 4D metamaterial lattice space has one spatial dimension, which is compactified. This concept has been demonstrated in our numerical simulations presented in Fig.4, which shows propagation of a Kaluza-Klein mode inside a lattice of waveguides, which emulates the Kaluza-Klein space-time described by eq.(2). The results in Fig. 4 were generated on the same lattice as Fig. 2a, but the software suite's periodic boundary conditions were used at the vertices to simulate the compactified "fifth dimension". In good agreement with the theory [9], an inset in Fig.4 clearly demonstrates a standing wave character of the Kaluza-Klein electromagnetic mode along the projected fifth lattice dimension (an individual waveguide in the inset is oriented along the projectred "fifth spatial dimension").

In conclusion, we have demonstrated that the metamaterial lattice-based approach to "optical space" design may supplement the more common approach of continuous engineering of $\varepsilon_{ik}$ and $\mu_{ik}$ tensors. We demonstrate this by presenting metamaterial lattice models of various hypercubic 4D spaces, which are

### References

[1] J. B. Pendry, D. Schurig, D.R. Smith, *Science* **312**, 1780 (2006).
[2] U. Leonhardt, *Science* **312**, 1777 (2006).
[3] U. Leonhardt and T. G. Philbin, *New J. Phys.* **8**, 247 (2006).
[4] X. Liu, C. Li, K. Yao, X. Meng, and F. Li, IEEE Antennas and Wireless Propagation Letters 8, 1154 (2009).
[5] P. Alitalo, F. Bongard, J.-F. Zürcher, J. Mosig, and S. Tretyakov, *Appl. Phys. Lett.* **94**, 014103 (2009).
[6] U. Leonhardt, and T. Tyc, *Science* **323**, 110 (2008).
[7] A. Louri, S. Furlonge, and C. Neocleous, *Appl. Optics* **35**, 6909 (1996).
[8] J. Gray, "Ideas of Space: Euclidean, Non-Euclidean, and Relativistic" (Clarendon Press, Oxford, 1989).
[9] I.I. Smolyaninov, *Phys. Rev. D* **65**, 047503 (2002).
[10] B. Edwards, A. Alu, M.E. Young, M. Silveirinha, and N. Engheta,, *Phys.Rev.Letters* **100**, 033903 (2008).
[11] I.I. Smolyaninov, *Journal of Optics* **13**, 024004 (2011). "Metamaterial "Multiverse"".

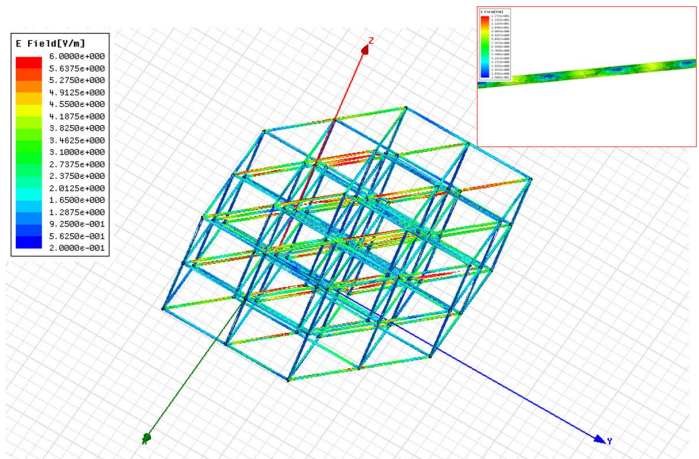

Fig. 4. (Color online) HFSS simulations of a Kaluza-Klein mode propagation inside a lattice of waveguides, which emulates the 5D spacetime described by eq.(2). The inset in the top right corner shows field magnitude inside an individual waveguide, which is oriented along the projected "fifth spatial dimension". Standing wave character of the Kaluza-Klein mode along the emulated fifth lattice dimension is clearly demonstrated.

impossible to emulate using conventional transformation optics. These lattice models provide a natural platform for various non-Euclidean broadband TO designs, as suggested in ref.[6]. In addition, studying fundamental linear and nonlinear optics in the emulated "4D space" may be quite an interesting fundamental exercise [11].